\documentclass[aps,prl,twocolumn,superscriptaddress,amsmath,amssymb,floatfix,preprintnumbers,nofootinbib,longbibliography]{revtex4-2}

\usepackage[normalem]{ulem}

\usepackage{graphicx} 
\usepackage{dcolumn} 
\usepackage{bm} 
\usepackage{hyperref} 
\usepackage{soul}
\usepackage{placeins}
\usepackage[export]{adjustbox}
\usepackage{ amssymb }
\usepackage{rotating}
\usepackage{siunitx}
\usepackage{lineno}
\usepackage{tabularx}
\usepackage[utf8]{inputenc}
\usepackage[T1]{fontenc}

\hypersetup{
    pdfnewwindow=true,    
    colorlinks=true,      
    linkcolor=blue,       
    citecolor=blue,       
    filecolor=blue,       
    urlcolor=blue         
}

\def\orcid#1{\href{https://orcid.org/#1}{\includegraphics[keepaspectratio,width=1em]{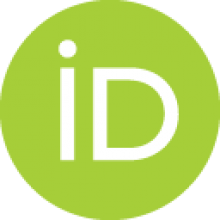}}}

\begin{document}
\title{Search for Signatures of Dark Matter Annihilation in the Galactic Center with HAWC}

\author{R.~Alfaro}\affiliation{Instituto de F\'{i}sica, Universidad Nacional Autónoma de México, Ciudad de Mexico, Mexico }

\author{C.~Alvarez}\affiliation{Universidad Autónoma de Chiapas, Tuxtla Gutiérrez, Chiapas, México}

\author{A.~Andrés}\affiliation{Instituto de Astronom\'{i}a, Universidad Nacional Autónoma de México, Ciudad de Mexico, Mexico }

\author{E.~Anita-Rangel}\affiliation{Instituto de Astronom\'{i}a, Universidad Nacional Autónoma de México, Ciudad de Mexico, Mexico }

\author{M.~Araya}\affiliation{Universidad de Costa Rica, San José 2060, Costa Rica}

\author{J.C.~Arteaga-Velázquez}\affiliation{Universidad Michoacana de San Nicolás de Hidalgo, Morelia, Mexico }

\author{D.~Avila Rojas}\affiliation{Instituto de Astronom\'{i}a, Universidad Nacional Autónoma de México, Ciudad de Mexico, Mexico }

\author{H.A.~Ayala Solares}\affiliation{Temple University, Department of Physics, 1925 N. 12th Street, Philadelphia, PA 19122, USA}

\author{R.~Babu}\affiliation{Department of Physics and Astronomy, Michigan State University, East Lansing, MI, USA }

\author{P.~Bangale}\affiliation{Temple University, Department of Physics, 1925 N. 12th Street, Philadelphia, PA 19122, USA}

% ...
\collaboration{HAWC Collaboration}
\hfill
\date{\today}

\begin{abstract}
We conduct an indirect dark matter (DM) search in the vicinity of the Galactic Center, focusing on a square region within $\pm 9^{\circ}$ in Galactic longitude and latutide, using 2,865 days of data ($\sim$8 years) from the High-Altitude Water Cherenkov (HAWC) Observatory. We explore DM particles within the Weakly Interacting Massive Particles framework with masses from 1 TeV to 10 PeV. Analyzing three annihilation channels ($b\bar{b}$, $\tau^{+}\tau^{-}$, $W^{+}W^{-}$) and three density profiles (Navarro-Frenk-White, Einasto, Burkert), we find no significant excess and set 95\% confidence-level upper limits on the velocity-weighted annihilation cross section. Our results provide the first constraints on DM particles well above 100 TeV using gamma-ray data from the vicinity of the Galactic Center, with the strongest limits $\mathcal{O}(10^{-24})$~cm$^{3}$/s, from the  $\tau^{+}\tau^{-}$ channel and the Einasto profile.

\end{abstract}

\maketitle


Astrophysical and cosmological evidence increasingly suggests that approximately 85\% of the Universe’s total matter density content consists of non-baryonic matter. This matter is commonly referred to as \textit{dark matter} due to its lack of electromagnetic interaction or luminosity~\cite{Feng:2010gw,Buckley:2017ijx,Patrignani_2016,Freese:2017idy}.
Beyond the matter component discussed above, the term ``Dark Universe" is used to collectively denote DM and dark energy, the latter introduced to account for the observed accelerated expansion of the Universe. The particle nature of DM remains an open question in modern physics. No suitable candidate exists within the Standard Model (SM). The proposed solutions range from relativistic thermal relic particles to condensates of weakly interacting non-relativistic light bosons~\citep{Duffy_2009,antypas2022new,PhysRevD.100.095018,KUSENKO20091}.

 One way of probing DM particles is to search for SM particles produced by DM self-annihilation. In this work, we focus on Weakly Interacting Massive Particles (WIMPs), which are accessible to astrophysical observations at the GeV–TeV scale~\citep{bertone2010particle,feng2010dark}. Many simple thermal WIMP models in the GeV–TeV range are now strongly constrained by a combination of null results from direct, indirect, and collider searches~\citep{Bergstrom2018, arcadi2025waning}. However, high-mass WIMPs remain viable candidates, providing open targets for indirect searches in the multi-TeV range, which is the focus of this work \citep{PhysRevD.102.083533, HAMBYE2020135553, PhysRevLett.126.081802, PhysRevD.101.115029}.

The Galactic Center (GC) is among the most promising targets for indirect DM detection due to the large DM density inferred from astrophysical observations~\citep{CESARINI2004267}. This region is expected to produce high gamma-ray fluxes from DM annihilation, making it a prime candidate for probing DM signals. Notably, an excess of GeV gamma rays observed by \textit{Fermi}-LAT from the GC has been interpreted by some as a potential signature of DM annihilation~\citep{DAYLAN20161,PhysRevLett.123.241101,murgia2020fermi},
    though other explanations such as millisecond pulsars cannot be ruled out~\citep{PhysRevLett.116.051102,bartels2018fermi,murgia2020fermi}
    These results further motivate the need to understand the GC emission at higher energies and testing scenarios of beyond SM physics. Previous searches for WIMP annihilation signatures in the GC and Galactic Halo have set constraints on the velocity-weighted cross section of DM particles with masses below 100~TeV~\citep{abdalla2022search,albert2023optimized,PhysRevD.108.102004,abazajian2020strong}. 

One of the main challenges for searches in this region is the complex astrophysical background, which includes emission from known gamma-ray sources, unresolved sources indistinguishable from the Galactic diffuse emission, and diffuse emission associated with the GC PeVatron~\citep{Albert_2024,hess2016acceleration}. However, instruments with sensitivities to gamma rays above 100 TeV are particularly powerful, since the Klein–Nishina suppression of inverse Compton scattering reduces leptonic background contributions and enhances sensitivity to potential DM signals~\citep{klein1928scattering}.

In this {\it Letter}, we use nearly eight years of data from the High-Altitude Water Cherenkov (HAWC) Gamma-Ray Observatory to search for DM particle masses between 1 TeV and 10 PeV, extending the previously reported range of DM masses~\citep{albert2023optimized} and constraining the velocity-weighted cross section of DM particles in the vicinity of the GC with masses well above 100~TeV for the first time. We consider a set of representative annihilation channels ($b\bar{b}$ quarks, $\tau^{+}\tau^{-}$ leptons, and $W^{+}W^{-}$ bosons) and spatial DM density profiles (Navarro-Frenk-White (NFW), Einasto, and Burkert). We mask known gamma-ray sources to avoid the contamination from their signal. No significant emission is found for any channel or density profile. Therefore, we set 95\% confidence-level (CL) upper limits on the velocity-weighted cross section of annihilation. In the following Sections, we describe the HAWC data and our analysis methods, discuss results, and then conclude. Further details are given in the Appendices \ref{app:profiles}--\ref{app:bur} of the Supplemental Material.

\emph{\textbf{HAWC Data.---}}
The HAWC Observatory, situated at 4,100 m altitude in Parque Nacional Pico de Orizaba, Mexico, detects cosmic and gamma rays with energies between hundreds of GeV and hundreds of TeV by capturing secondary particles produced in atmospheric air showers initiated by astrophysical primaries~\cite{2017ApJ...843...39A}. Its array of 300 water-Cherenkov detectors operates nearly continuously~\citep{ABEYSEKARA2023168253}. The vast majority of showers detected originate from hadronic cosmic rays, creating an isotropic and steady background. Using shower topology cuts~\cite{Albert_2024per}, we preserve gamma-ray signals while filtering out broader, clumpier hadronic showers. In 2023, the event-reconstruction algorithms were greatly improved (``Pass 5'' version), extending the maximum zenith angle to $\sim$56$^{\circ}$, allowing better sensitivity at the edges of HAWC's field of view~\cite{Albert_2024per}. Thanks to these enhancements, HAWC has
unique sensitivity to the GC at energies above 40 TeV~\cite{Albert_2024}, allowing us to probe uncharted DM particle masses. Here we use 2,865 days of data and improved reconstruction algorithms to directly target the region surrounding the GC ($\pm 9^{\circ}$). 

\emph{\textbf{Methods.---}}
For a given DM source and a stable SM particle, the flux of SM particles from DM annihilation into gamma rays ($\gamma$) is described as follows:

\begin{equation}\label{eq:id_dm_flux}
  \dfrac{d\Phi_\gamma}{dE_{\gamma}} = \frac{\langle \sigma v \rangle}{8 \pi M^2_{\chi}} \frac{dN_\gamma}{dE_\gamma} \times \int_{\text{source}}{d\Omega \int_{\text{l.o.s.}}{\rho^2_\chi dl(r,\theta')}}\,,
\end{equation}
where $\langle \sigma v \rangle$ represents the velocity-weighted annihilation cross section of DM to SM particles and $M_\chi$ is the DM particle mass. The term $\frac{dN_{\gamma}}{dE_\gamma}$ denotes the number of photons produced per annihilation. The integral is performed over the solid angle, $d\Omega$, and the line of sight (l.o.s.). The term $\rho_{\chi}$ represents the DM density at a given location, $(r, \theta’)$, in the sky.

Various possibilities can be considered for the distribution of DM in the Milky Way. The NFW profile, with a cusp at the GC, is a traditional benchmark from $N$-body simulations~\citep{1996ApJ...462..563N}. The Einasto profile is broader at kiloparsec scales and favored by recent simulations~\citep{Merritt_2006,10.1111/j.1365-2966.2009.15878.x}. The Burkert profile~\citep{Burkert_1995}, which features a constant-density core, aligns with observed Galactic rotation curves but is in tension with cold DM-only simulations that predict a steep central cusp. In order to include the spatial signature hypotheses in the likelihood analysis, we calculate the astrophysical component of Equation~\ref{eq:id_dm_flux}, referred to as the $J$-factor:
\begin{equation}
  J\text{-factor} = \int_{\text{source}}{d\Omega \int_{\text{l.o.s}}{\rho^2_\chi dl(r,\theta')}}\,,
\end{equation}
for each pixel in a map of the region of interest. These $J$-factor spatial distributions are produced using \texttt{Gammapy}~\citep{gammapy:2023}--- an open-source Python package for gamma-ray astronomy. More details can be found in Appendix \ref{app:profiles}.


The spectral models implemented in this analysis are imported from \texttt{HDMSpectra}~\citep{Bauer_2021}, which provides the expected photon spectra for DM annihilation to different SM channels, constituting a wide variety of hypotheses. These annihilation spectra assume prompt emission, neglecting any subsequent propagation or energy loss effects~\citep{Porter_2022}. Gamma rays with TeV energies also undergo attenuation by interaction with the cosmic microwave background and other galactic radiation fields~\citep{2015JCAP...10..014E}. To estimate the potential impact on high-mass DM scenarios, we conservatively apply a uniform 50\% attenuation to fluxes above 200~TeV, approximating the trend shown in Fig.~2 of Ref.~\citenum{Zhang:2024xkh}. This provides an upper bound on the effect of absorption for PeV-scale DM, neglecting possible enhancement from secondary photons produced by down-scattering of higher-energy gamma rays \citep{PhysRevD.110.103021}. Consequently, the 95\% CL limits for the highest masses are weakened by $\sim$10--40\%. HAWC’s sensitivity peaks around 100~TeV, below the regime where attenuation becomes extreme, so the most constraining part of the spectrum is largely unaffected. The limits below 200~TeV remain conservative, and any secondary photon contribution could in fact increase the signal up to a factor of 2~\citep{Zhang:2024xkh}. Any resulting change in limits would still be contained within the statistical uncertainties of expected limits as shown in Fig. \ref{fig:ts}. To include the spatial and energy response of the detector, the predicted gamma-ray signal is convolved with the energy-dependent point spread function and the detector effective area for each event \citep{Albert_2024per}. This ensures that the likelihood analysis properly accounts for the finite angular and energy resolution of HAWC. The Test Statistic (TS) is computed with the log-likelihood ratio test, which compares the expected number of events from DM annihilation to the background-only hypothesis. The best-fit signal parameters correspond to the value of $\langle \sigma v \rangle$ (Equation~\ref{eq:id_dm_flux}) that maximizes the likelihood.
\begin{center}
\begin{figure}[h] 
\centering
{\includegraphics[width=0.4\textwidth]{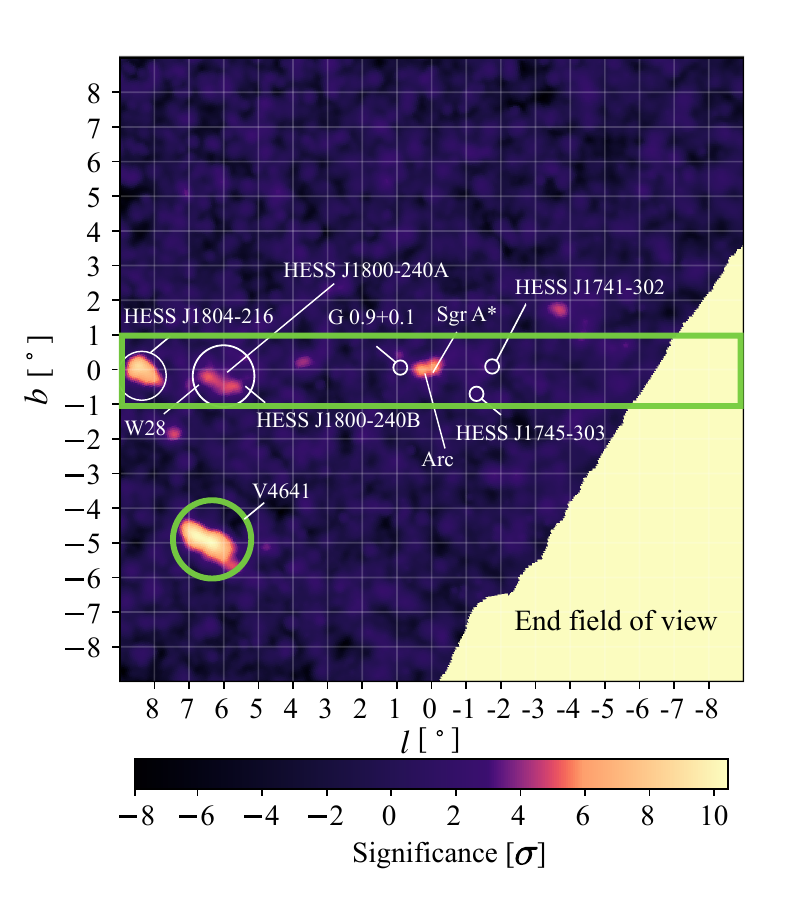}}
\caption{Significance map of the region of interest used for this analysis. The significance is defined as $\sqrt{\text{TS}}$. The labels indicate very-high-energy gamma-ray sources. The green rectangle and circle outline the masks. The bottom right region is the end of HAWC's field of view ($\sim$56$^{\circ}$ zenith angle) and is excluded from the likelihood fit. Color available online.\label{fig:maskslabel} }
\end{figure}
\end{center}

\begin{figure*}[t]
\makebox[0.8\width][c]{
\begin{tabular}{@{}cc@{}}
\includegraphics[width=0.4\textwidth]{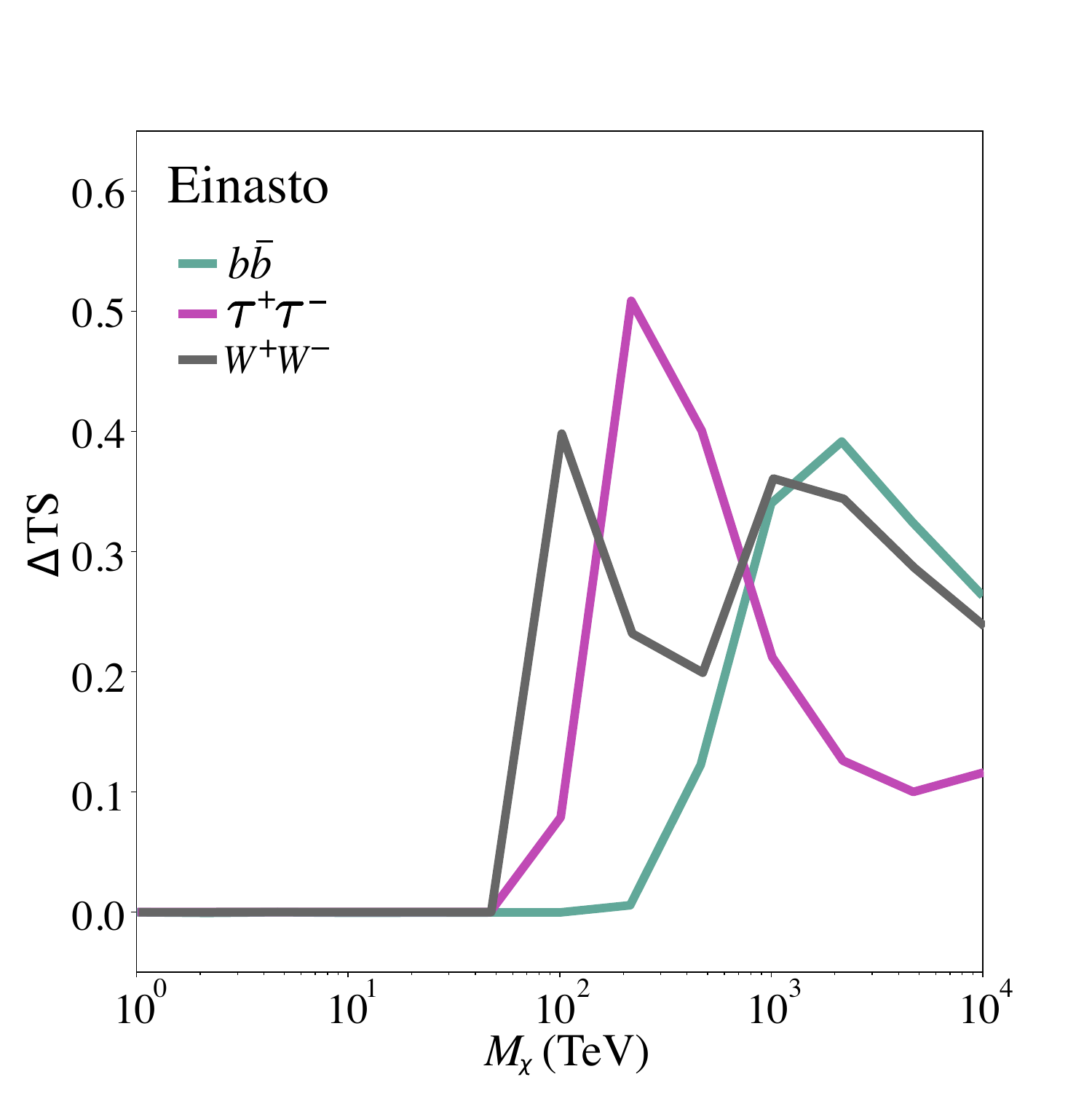} &
\includegraphics[width=0.4\textwidth]{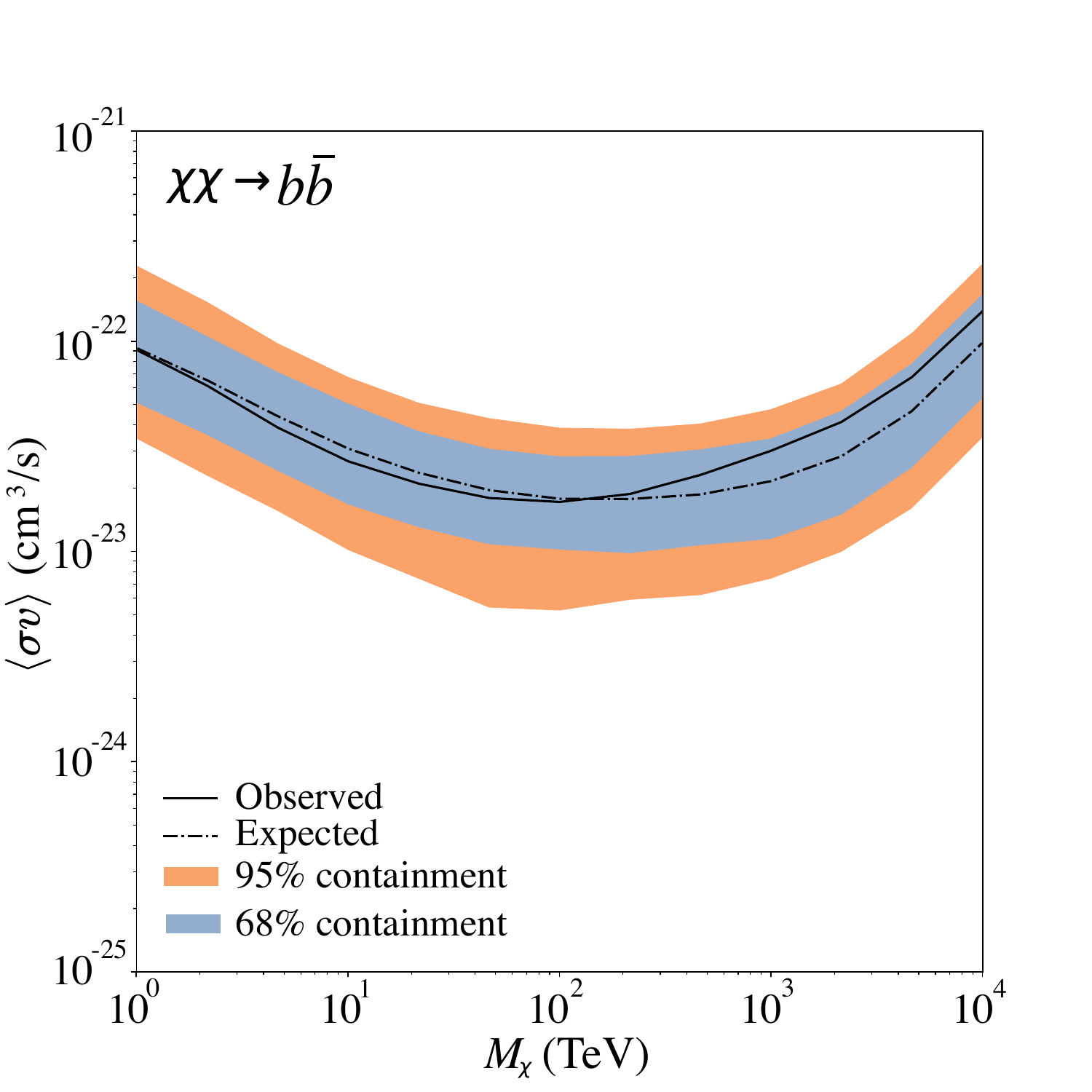} \\
\includegraphics[width=0.4\textwidth]{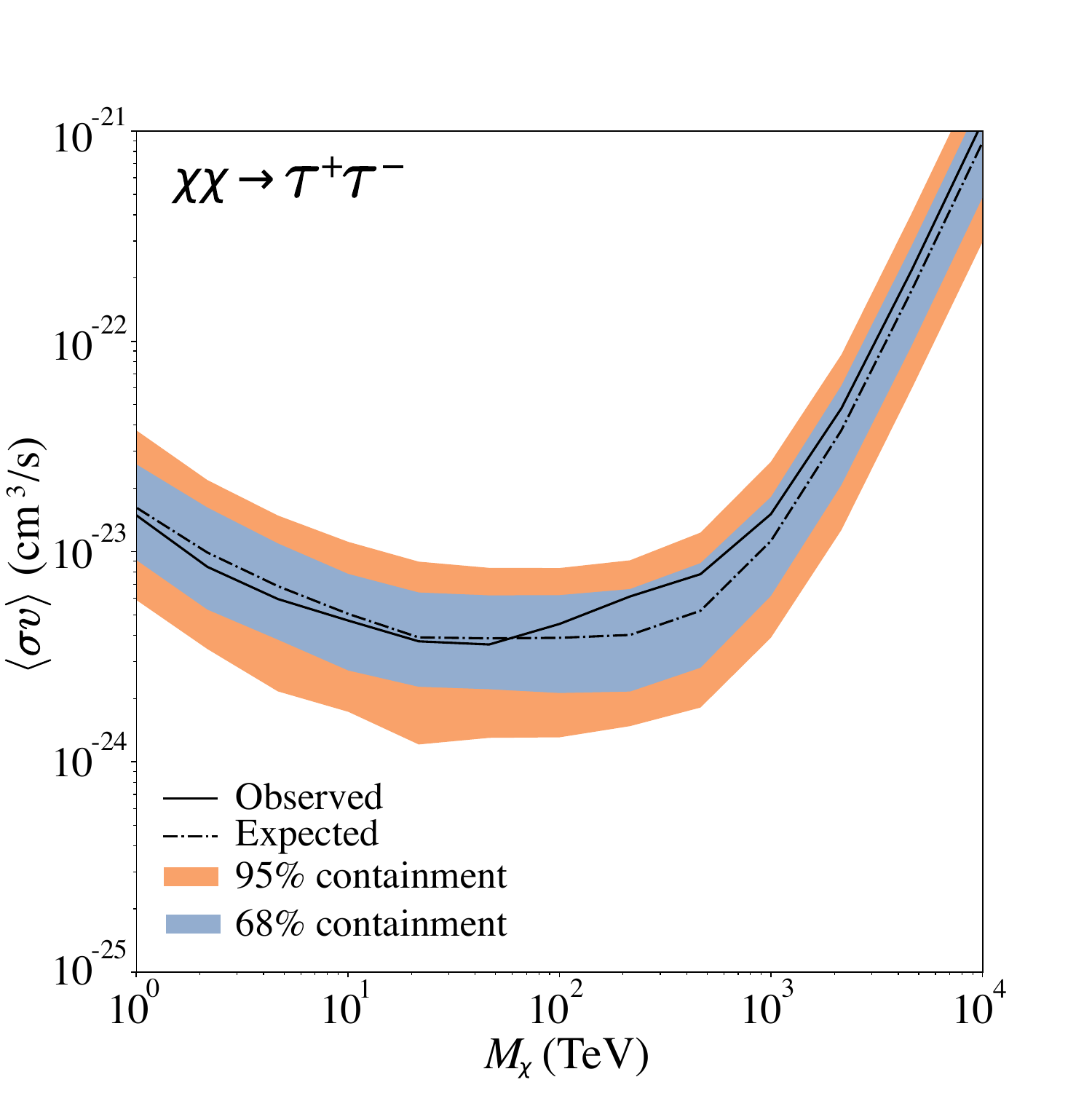} &
\includegraphics[width=0.4\textwidth]{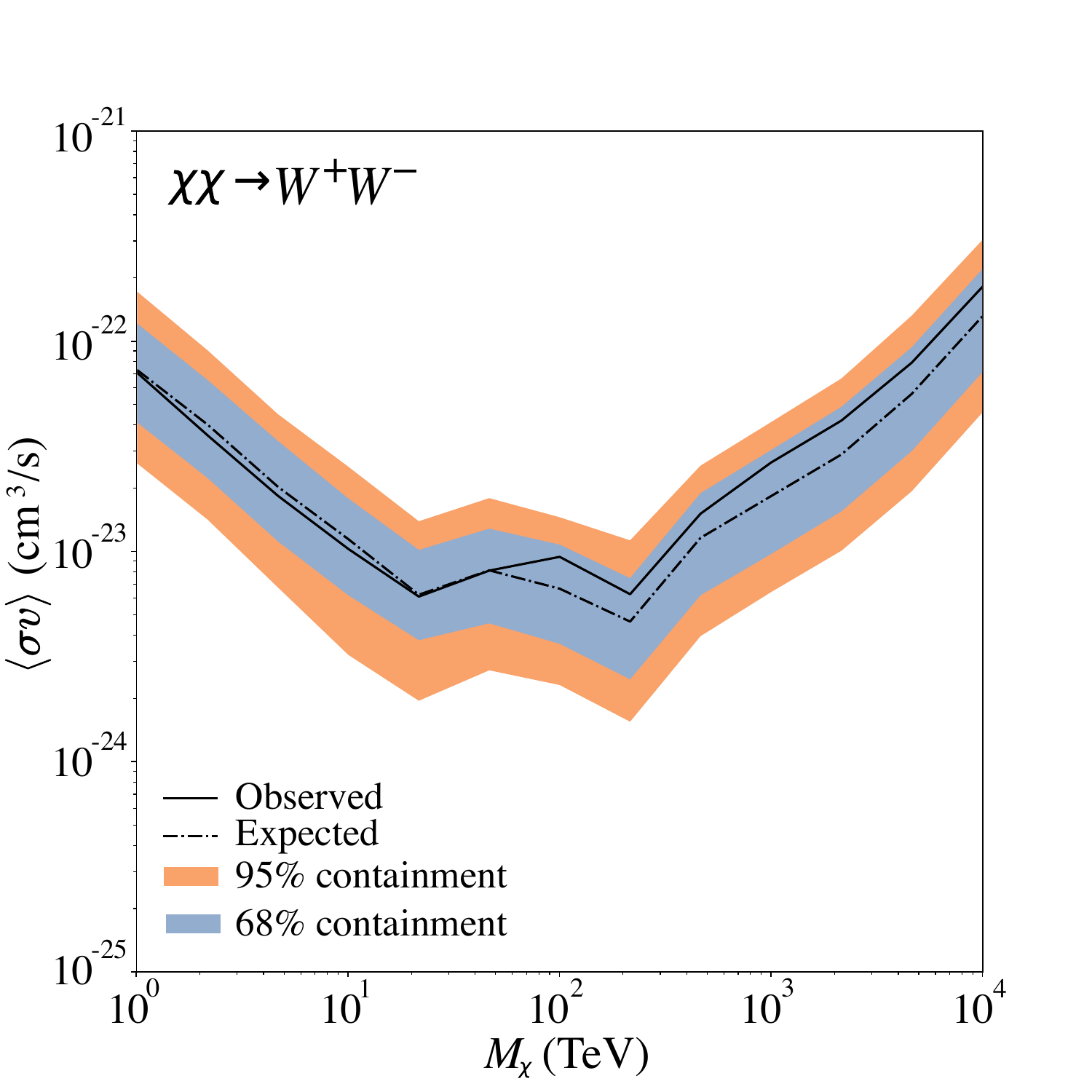} \\
\end{tabular}
}
\caption{Results assuming an Einasto density profile. \textbf{Upper left}: best-fit $\Delta$TS for each annihilation channel (background and signal vs background-only model). \textbf{Upper right},  \textbf{lower left}, \textbf{lower right}: containment bands for the 1 and 2$\sigma$ expected limits compared to the observed upper limit at the 95\% CL for the channels: $b\bar{b}$, $\tau^{+}\tau^{-}$, and $W^{+}W^{-}$, respectively. Color version online. \label{fig:ts}}
\end{figure*}

 In Figure \ref{fig:maskslabel}, we show the significance map of the $\pm 9 ^{\circ}$ region around the GC, where the analysis is conducted. The blank region in the bottom right corresponds to zenith angles beyond $\sim$56$^\circ$, where HAWC’s reconstruction and simulations are not optimized. This region is therefore excluded from the analysis.

To mitigate contamination from known gamma-ray sources, we mask them in the HAWC data by setting the $J$-factor in the corresponding pixels to zero. The Galactic Plane is masked with a $\pm1^{\circ}$ band, effectively removing sources with spectral index $<2.5$ or coincident with $>3\sigma$ hotspots. The only known source outside this band, V4641 Sagittarius \cite{alfaro2024ultra}, is masked separately.  The size of these masks are conservative, with sizes larger than HAWC’s angular resolution at this declination.

\emph{\textbf{Results.---}} In Figure \ref{fig:ts}, we show the best-fit $\Delta$TS for each annihilation channel, assuming an Einasto density profile (see  Appendix~\ref{app:bur} for NFW and Burkert). No significant excess was found in the probed mass range (1~TeV--10~PeV). Therefore, upper limits  on $\langle \sigma v \rangle$ were placed at the 95\% CL (see details in Appendix~\ref{app:ul}). 

We also show in Figure \ref{fig:ts} the upper limit on $\langle \sigma v \rangle$ at the 95\% CL for the three channels assuming the Einasto density profile. The mean expected upper limit is also shown along with the 68\% and 95\% containment bands, which are obtained from 300 Poisson realizations of the background (more details in Appendix~\ref{app:el}). In all three cases, the limits deviate from expectations only within the 95\% confidence band. This suggests that the deviations are driven purely by Poisson fluctuations rather than an unaccounted source or a mischaracterization of the background. The $\tau^+\tau^-$ channel yields the strongest limits due to its hard gamma-ray spectrum, to which HAWC is more sensitive. Detector systematic uncertainties in HAWC are evaluated by performing the analysis with modified detector response functions. In all cases, the resulting uncertainties are found to be below 30\% (more details in Appendix~\ref{app:syst}).

In Figure~\ref{fig:braziltautau}, we compare our limits with previous results from HAWC, the High-Energy Stereoscopic System (H.E.S.S.), and IceCube. Compared to HAWC’s 2023 Galactic Halo analysis~\citep{albert2023optimized}, we extend the mass range by a factor of four using more data and improved reconstruction. Our limits are up to twice as strong between 10–100~TeV for $b\bar{b}$ and $\tau^{+}\tau^{-}$ and are comparable for $W^+W^-$.

H.E.S.S. has greater sensitivity below 70~TeV due to its finer angular resolution and optimal location in the Southern Hemisphere, while HAWC—observing the GC at high zenith—probes higher masses, beyond H.E.S.S.’s reach. IceCube’s GC limits are stronger below 7~TeV, but our limits become comparable or stronger above that, particularly for $b\bar{b}$. An equivalent figure using the Burkert profile can be found in Appendix \ref{app:bur} of the Supplemental Material.

\begin{center}
\begin{figure}[ht!]
\centering
{\includegraphics[width=0.45\textwidth]{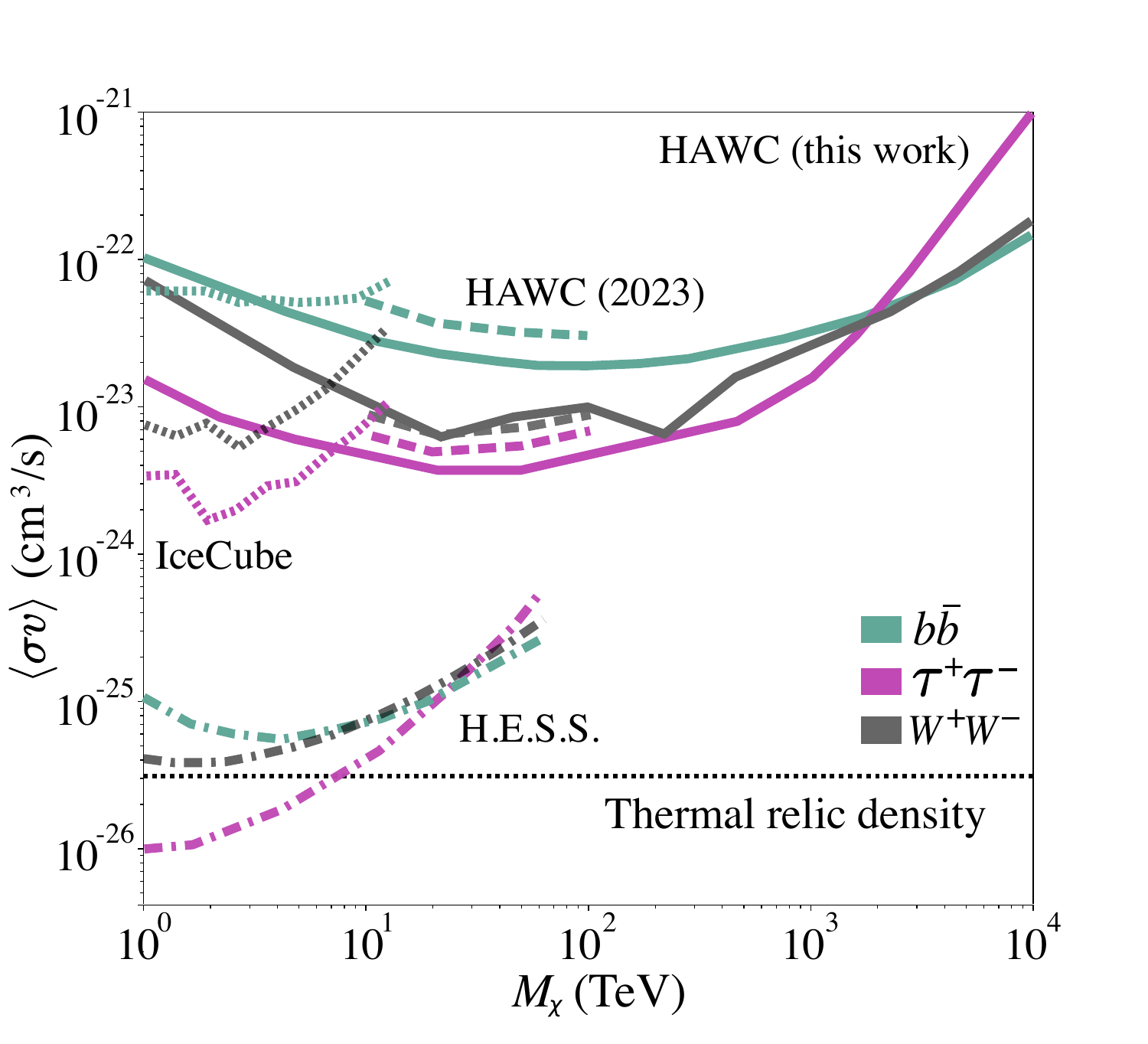}}
\caption{Comparison of all channels. The IceCube limits (dotted lines), originally derived assuming an NFW profile~\citep{PhysRevD.108.102004}, have been rescaled to Einasto by a factor of 0.45. Also shown are HAWC (``Pass~4’’; dashed lines)~\citep{albert2023optimized} and H.E.S.S. limits~\citep{abdalla2022search} (dash-dotted lines). Color version online. \label{fig:braziltautau}}
\end{figure}
\end{center}

\emph{\textbf{Summary.---}}
In this {\it Letter}, we present an indirect DM search in the vicinity of the GC using nearly eight years of HAWC data, probing WIMP annihilation in the 1~TeV--10~PeV mass range. We masked the Galactic Plane and other known sources to mitigate gamma-ray contamination. The lack of a significant excess in all density profiles and channels probed allows us to place 95\% CL constraints on the velocity-weighted annihilation cross section of WIMPs. 

These results provide the first gamma-ray constraints above 100~TeV from the vicinity of the GC. The strongest limits are found for the $\tau^+\tau^-$ channel assuming an Einasto profile. While the maximum mass of thermal freeze-out WIMPs is subject to the unitarity bound of about 130 TeV, our results are applicable to various models that predict non-thermal mechanisms of DM production in which the masses can be as high as 10$^6$ TeV~\citep{Kramer:2020sbb,PhysRevD.109.035018}. Although H.E.S.S. is more sensitive at lower energies, HAWC uniquely probes the multi-TeV to PeV mass range. Compared to IceCube and prior HAWC results, our limits improve significantly above 10~TeV. Systematic uncertainties remain below 30\% for all profiles and channels. While uncertainties related to the $J$-factor and radiative corrections to the channel spectra are not explicitly included, the differences between the assumed density profiles encompass the dominant uncertainties in the DM distribution~\citep{Cirelli:2010xx}. 


This study extends the experimental frontier of indirect DM detection to PeV-scale masses and strengthens constraints on particle annihilation in one of the most promising regions for DM concentration. Our results help narrow the parameter space of heavy thermal relics, while providing valuable input for both particle theory and GC astrophysics.

\bigskip
\begin{acknowledgements}

{\bf\emph{Acknowledgments.---}}
We acknowledge the support from: the US National Science Foundation (NSF); the US Department of Energy Office of High-Energy Physics; the Laboratory Directed Research and Development (LDRD) program of Los Alamos National Laboratory; Consejo Nacional de Ciencia y Tecnolog\'{i}a (CONACyT), M\'{e}xico, grants LNC-2023-117, 271051, 232656, 260378, 179588, 254964, 258865, 243290, 132197, A1-S-46288, A1-S-22784, CF-2023-I-645, c\'{a}tedras 873, 1563, 341, 323, Red HAWC, M\'{e}xico; DGAPA-UNAM grants IG101323, IN111716-3, IN111419, IA102019, IN106521, IN114924, IN110521 , IN102223; VIEP-BUAP; PIFI 2012, 2013, PROFOCIE 2014, 2015; the University of Wisconsin Alumni Research Foundation; the Institute of Geophysics, Planetary Physics, and Signatures at Los Alamos National Laboratory; Polish Science Centre grant, 2024/53/B/ST9/02671; Coordinaci\'{o}n de la Investigaci\'{o}n Cient\'{i}fica de la Universidad Michoacana; Royal Society - Newton Advanced Fellowship 180385; Gobierno de España and European Union-NextGenerationEU, grant CNS2023- 144099; The Program Management Unit for Human Resources \& Institutional Development, Research and Innovation, NXPO (grant number B16F630069); Coordinaci\'{o}n General Acad\'{e}mica e Innovaci\'{o}n (CGAI-UdeG), PRODEP-SEP UDG-CA-499; Institute of Cosmic Ray Research (ICRR), University of Tokyo. H.F. acknowledges support by NASA under award number 80GSFC21M0002. C.R. acknowledges support from National Research Foundation of Korea (RS-2023-00280210). We also acknowledge the significant contributions over many years of Stefan Westerhoff, Gaurang Yodh and Arnulfo Zepeda Dom\'inguez, all deceased members of the HAWC collaboration. Thanks to Scott Delay, Luciano D\'{i}az and Eduardo Murrieta for technical support.

\end{acknowledgements}



\bibliographystyle{apsrev4-1}
\bibliography{mbib}


\clearpage

\onecolumngrid
\appendix

\ifx \standalonesupplemental\undefined
\setcounter{page}{1}
\setcounter{figure}{0}
\setcounter{table}{0}
\setcounter{equation}{0}
\setcounter{secnumdepth}{2}
\fi

\renewcommand{\thepage}{Supplemental Material -- S\arabic{page}}
\renewcommand{\theequation}{S\arabic{equation}}
\renewcommand{\figurename}{SUPPL. FIG.}
\renewcommand{\tablename}{SUPPL. TABLE}

\clearpage
\newpage

\centerline{\Large {\bf Supplemental Material for}}
\medskip

\centerline{\Large \emph{Search for Signatures of Dark Matter Annihilation in the Galactic Center with HAWC}}
\medskip

\centerline{\large HAWC Collaboration}
\bigskip

We provide additional details in the appendices below to aid the reading of the paper. Appendix \ref{app:profiles} gives details regarding the density profiles and $J$-factor computation. Appendices \ref{app:ul} and \ref{app:el} describe how the upper and expected limits were computed. Appendix \ref{app:syst} gives details about the calculation of systematic uncertainties. Finally, Appendix \ref{app:bur} shows the best-fit TS results for NFW and Burkert, as well as the limits for the Burkert density profile.

\section{Density profiles for the Galactic Center and $J$-factor computation in the regions of interest}\label{app:profiles}


Various possibilities can be considered for the distribution of DM in the Milky Way. Here we focus on three widely used profiles—NFW, Einasto, and Burkert—shown in Supplemental Figure~\ref{fig:profiles}. Although these do not span the full parameter space of possible halo models, they provide a representative set for evaluating how assumptions about the density profile affect DM signal predictions. The functional forms of these profiles are given by:

\begin{align}
    \rho_{\text{NFW}}(r)&=\rho_{s}\dfrac{r_{s}}{r}\left( 1+\dfrac{r}{r_{s}}\right)^{-2}\,,\\
    \rho_{\text{Ein}}(r)&=\rho_{s}\exp\left[-\dfrac{2}{\alpha}\left(\dfrac{r}{r_{s}} \right)^{\alpha}-1 \right]\,,\\
    \rho_{\text{Bur}}(r)&=\dfrac{\rho_{s}}{(1+r/r_{s})(1+(r/r_{s})^{2})}\,,
\end{align}
where $\rho_{s}=0.39$ GeV/cm$^{3}$ is the local density, $\alpha=$ 0.17 is an empirical fit to numerical simulations of DM halos, and $r_{s}=20$ kpc is a scale value based on theoretical models and observational constraints \citep{posti2019mass,iocco2015evidence}.

\begin{center}
\begin{figure}[ht!]
\centering
{\includegraphics[width=0.5\textwidth]{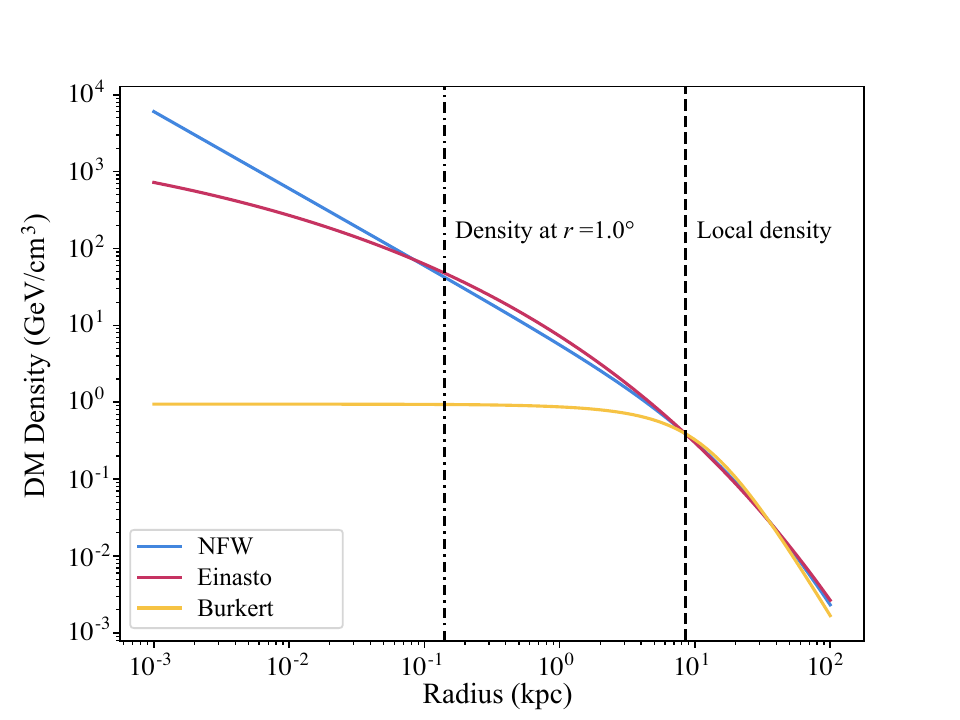}}
\caption{The three DM density profiles considered in this analysis: NFW (blue), Einasto (red), and Burkert (yellow). It is worth noting that the Burkert profile features a significantly more flat, low-density center (``cored'') DM distribution, resulting in lower central densities at the GC compared to the cusped NFW and Einasto profiles. The dash-dotted line marks the distance equivalent to the width of the Galactic plane mask (1.0$^{\circ}$). The dashed line shows the distance from Earth to the GC. \label{fig:profiles}}
\end{figure}
\end{center}

\section{Computation of upper limits}\label{app:ul}
\setcounter{equation}{3}

The TS is computed with the log-likelihood ratio test:
\begin{equation}\label{eq:ts}
    \text{TS} = -2 \, \ln \left( \frac{L_{\text{null}}}{L_{\text{model}}} \right)\,,
\end{equation}  
where  $L_{\text{model}}$  is the likelihood of the model with a DM annihilation source and  $L_{\text{null}}$  is the likelihood of the background-only alternative. The best-fit signal parameters are found where $\langle \sigma v \rangle$ in Equation~\ref{eq:id_dm_flux} maximizes the likelihood.

To determine the upper limit at the 96\% CL, we began by calculating the TS for $\langle \sigma v \rangle$. The value of $\langle \sigma v \rangle$ that yielded the largest TS value from Equation \ref{eq:ts}, TS$_{\text{max}}$, was then determined by scanning over increasing values of $\langle \sigma v \rangle$. To find the 96\% CL limit, we scan $\langle \sigma v \rangle$ until the TS decreases by $\Delta \text{TS}=2.71$, which is determined from the appropriate distribution for a non-negative signal parameter \citep{cowan2011asymptotic} and satisfies:

\begin{equation}\label{eq:confint_ul}
\frac{1}{2} + \frac{1}{2}\,F_{\chi^2}\!\big(\Delta \text{TS}\big) = 0.95,
\end{equation}
where $F_{\chi^2}$ is the cumulative distribution function of a $\chi^2$ distribution with 1 degree of freedom.  

The upper limit on the velocity-weighted cross section, $\langle \sigma v \rangle_{95}$, is then defined as the value for which
\[
\text{TS}(\langle \sigma v \rangle_{95}) = \text{TS}_{\text{max}} - \Delta \text{TS}.
\]
This procedure is repeated for each DM density profile and annihilation channel to obtain the 95\% CL upper limits.

\section{Computation of expected limits and exclusion bands}\label{app:el}

 The expected limits and their statistical variations are computed under the background-only hypothesis, which is based on HAWC’s direct-integration background model. This model provides the expected background counts in each analysis bin by averaging event rates over long time windows in local coordinates \citep{2012ApJ...750...63A, Milagro:2003yym}, capturing exposure variations and large-scale anisotropies.
 
Pseudo-datasets are generated by replacing the background counts in each bin with a random draw from a Poisson distribution using the direct-integration value as its mean. This procedure models statistical fluctuations around the expected background and produces a consistent, validated construction of the expected TS distribution. Sideband or off-source regions are not used, as their detector response, exposure, and background morphology may differ from the target region and could bias the resulting distribution.

Using 300 such Poisson realizations of the null hypothesis, we calculate the 68\% and 95\% containment bands for the expected limits.

\section{Systematic uncertainties}\label{app:syst}


Systematic uncertainties in this analysis capture variations in the detector response that affect both the spatial and energy characteristics of the predicted DM signal. The signal is convolved with HAWC’s energy-dependent point spread function and effective area, and uncertainties in these detector properties propagate through this convolution. Since we are studying the same region as in Reference \citenum{Albert_2024}, we selected the two configurations of the detector that resulted in the largest systematic uncertainty in that analysis, late-light simulation and absolute PMT efficiency comparing different epochs in time, which are described in Reference \citenum{Albert_2024per}. The systematic uncertainties are $<$30\% for all three density profiles.

\section{NFW and Burkert profile}\label{app:bur}

The best-fit $\Delta$TS vs mass plots are shown in Supplemental  Figure \ref{fig:tsbur} assuming the NFW and Burkert density profiles. 

In Supplemental Figure \ref{fig:burkert}, we show the upper limits derived using the Burkert profile. These are approximately two orders of magnitude weaker than those obtained assuming the Einasto and NFW profiles. The absence of a central cusp results in a lower predicted gamma-ray flux, necessitating a larger region of interest to capture comparable emission. However, in this study we maintained a consistent region of interest across different profiles restricting the detectable flux, thereby reducing sensitivity and weakening the constraints relative to the cuspy NFW and Einasto profiles. The same argument can be used to explain the difference between our limits and the ones obtained by HAWC targeting the Galactic Halo instead.

\begin{figure*}[t]
\makebox[0.8\width][c]{
\begin{tabular}{@{}cc@{}}
\includegraphics[width=0.4\textwidth]{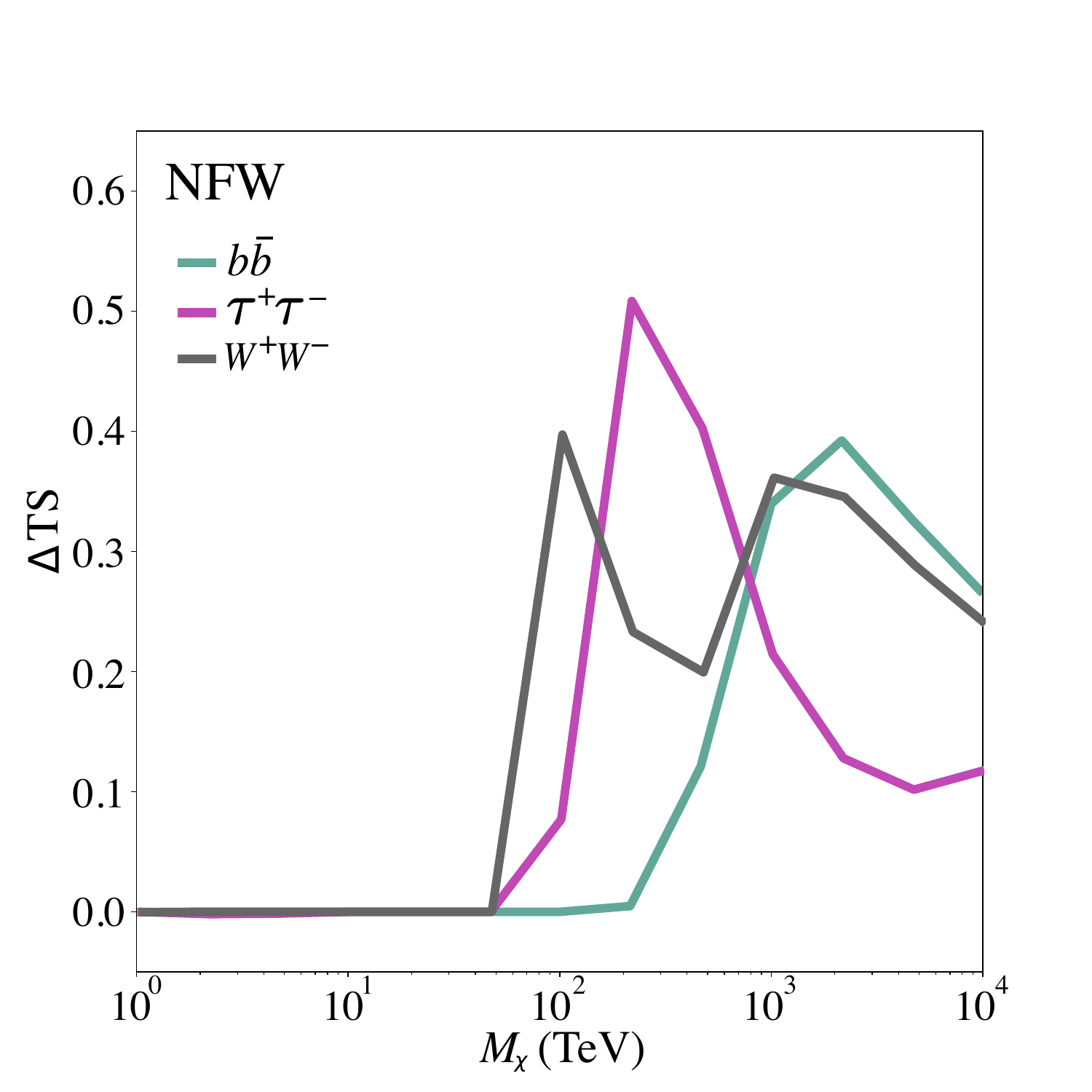} &
\includegraphics[width=0.4\textwidth]{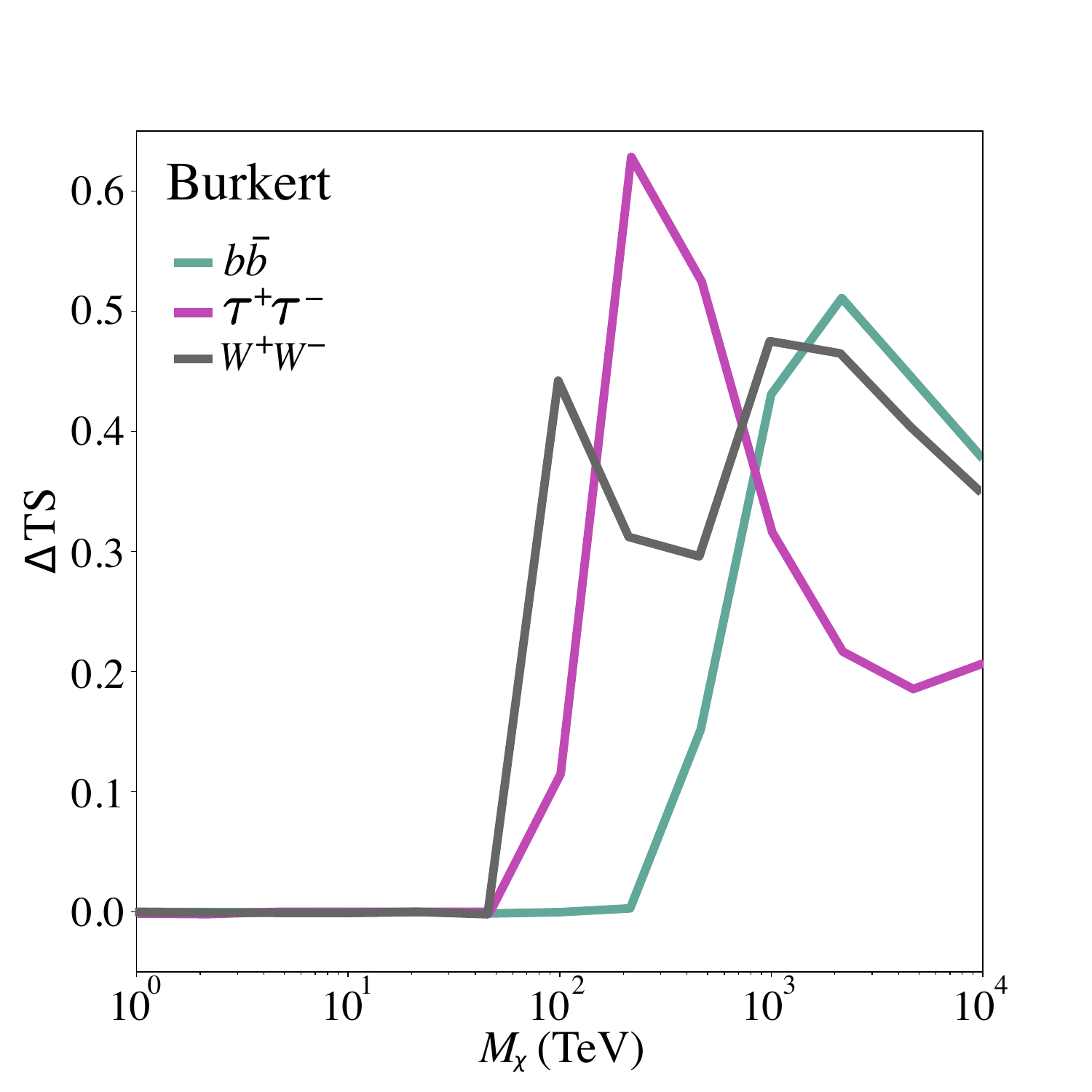} \\
\end{tabular}
}
\caption{Best-fit $\Delta$TS for each annihilation channel (background and signal vs background-only model) assuming the \textbf{left} NFW and \textbf{right} Burkert density profiles. \label{fig:tsbur}}
\end{figure*}

\begin{center}
\begin{figure}[ht!]
\centering
{\includegraphics[width=0.40\textwidth]{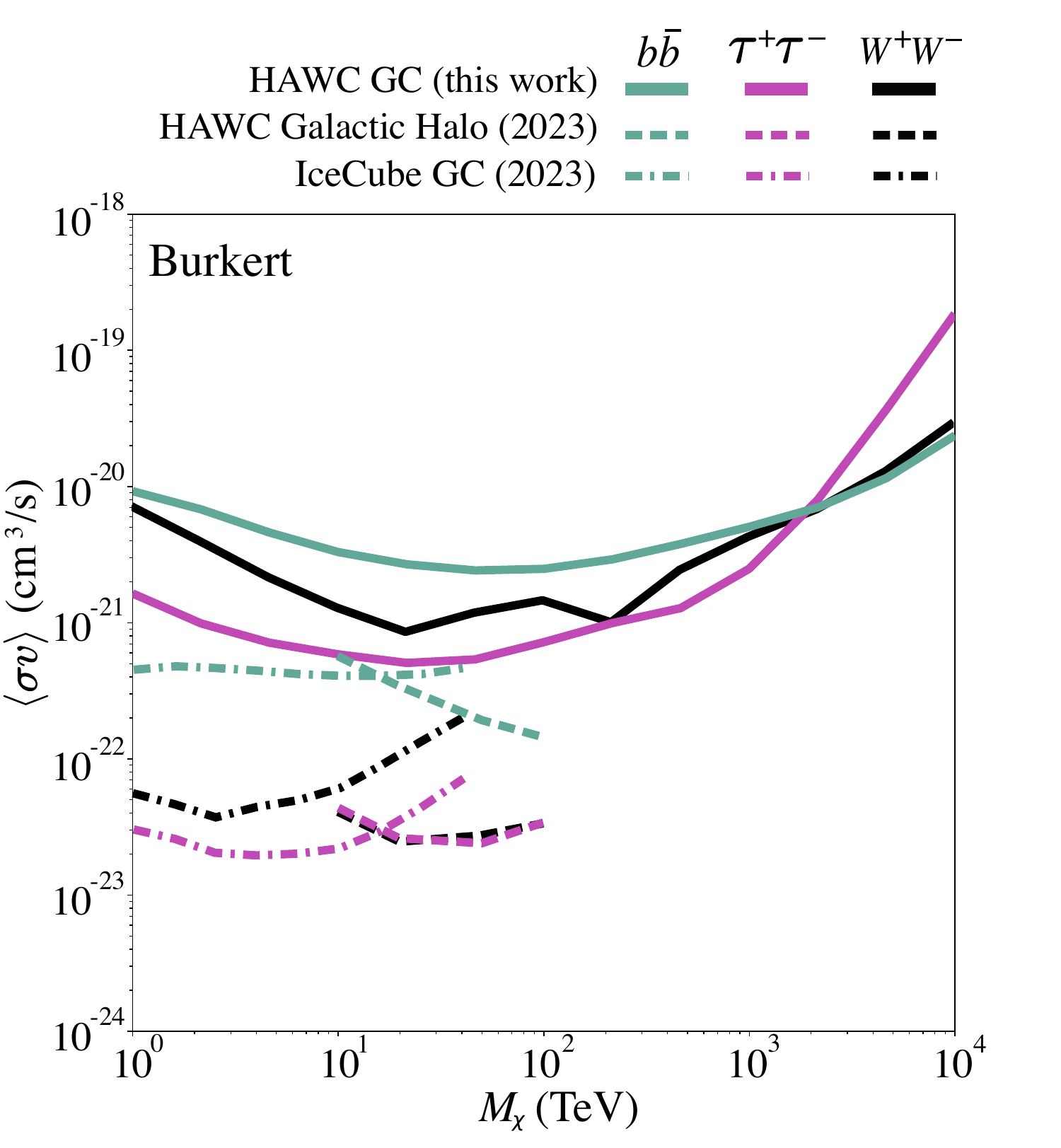}}
\caption{Upper limits at 95\% CL on the velocity-weighted annihilation cross section for the channels: $b\bar{b}$ (cyan), $\tau^{+}\tau^{-}$ (purple), and $W^{+}W^{-}$ (black) assuming the Burkert profile compared to HAWC (``Pass 4'' version; dashed lines) \citep{albert2023optimized} and IceCube \citep{PhysRevD.108.102004} (dotted lines). Note that the $y$-axis is different from the one in Figure \ref{fig:braziltautau} (bottom right). Color available online. \label{fig:burkert} }
\end{figure}
\end{center}


\end{document}